\RequirePackage{fix-cm}
\documentclass[natbib,twocolumn,final]{svjour3}          
\usepackage{soul}
\usepackage{amssymb}
\usepackage{hyperref}
\usepackage{mathtools, cuted}
\usepackage{commath}
\usepackage{lineno}
\usepackage{siunitx}
\usepackage{graphicx,amssymb,amsmath}
\usepackage{txfonts}
\usepackage{natbib}
\usepackage{mathptmx}      
\usepackage{aps-bibstyle}  
\usepackage[first=0,last=9]{lcg}
\usepackage[table]{xcolor}
 \journalname{to be inserted}
\newcommand{\aap}{{Astron. Astrophys.}}

\begin{document}
\title{The analysis of restricted five--body problem within frame of variable mass}
\author{Md Sanam Suraj         \and
        Elbaz I. Abouelmagd \and
        Rajiv Aggarwal\and
       Amit Mittal
}
\institute{Md Sanam Suraj \at
    Department of Mathematics,
    Sri Aurobindo College, University of Delhi,  New Delhi-110017, Delhi, India\\
    \email{\url{mdsanamsuraj@gmail.com}}\\
    \email{\url{mdsanamsuraj@aurobindo.du.ac.in}}           
\and
Elbaz I. Abouelmagd\at
Nonlinear Analysis and Applied Mathematics Research Group (NAAM),
 Department of Mathematics, Faculty of Science, King Abdulaziz University, Jeddah, Saudi Arabia.\\
 Celestial Mechanics Unit, Astronomy Department,
National Research Institute of Astronomy and Geophysics (NRIAG),
Helwan--11421, Cairo, Egypt.\\
 \email{\url{elbaz.abouelmagd@nriag.sci.eg or eabouelmagd@gmail.com}}
\and
Amit Mittal\at
ARSD College, University of Delhi, New Delhi-110021, Delhi,  India\\
 \email{\url{to.amitmittal@gmail.com}}
           \and
  Rajiv Aggarwal \at
  Department of Mathematics,
  Deshbandhu College, University of Delhi, New Delhi-110019, Delhi, India\\
              \email{\url{rajiv_agg1973@yahoo.com}}
}

\date{Received: date / Accepted: date}
\maketitle
\begin{abstract}
In the framework of restricted five bodies problem, the existence
and stability of the libration points are explored and analysed numerically,
under the effect of non--isotropic mass variation of the fifth body
(test particle or infinitesimal body). The evolution of the positions of these
points and the possible regions of motion are illustrated, as a function
of the perturbation parameter.
We perform a systematic investigation in an attempt to understand how the
perturbation parameter due to variable mass of the fifth body, affects the
positions, movement and stability of the libration points.
In addition, we have revealed how the domain of the basins of convergence
associated with the libration points are substantially influenced by
the perturbation parameter.
\end{abstract}
\section{Introduction}
\label{intro}
The $N-$body problem is not only fascinating but also presents an interesting
challenge to the researchers and scientists. In general the space missions
could be designed within frame of the $N-$body problem, which can be reduced to
three, four or five--body problem in some cases, etc. The dynamical system of
restricted five--body problem
has a great significance in celestial mechanics. So many researchers over
the world are currently interested in studying and solving aforesaid problem, i.e.,
the restricted five--body problem with various perturbation forces.
The restricted five--body problem primarily takes into account a fifth body referred
as the test particle with negligible mass, which does not influence the motion of
four primaries moving in circular orbits around their common center of mass.
This problem is a simple extension of four--body problem.
Some of work are available on the planar central configuration of
$N-$bodies with $N=4, 5$ and $ 7$,
see for details \cite{esgl}, \cite{PK07}, \cite{jaume}.

The history of restricted problem of $N-$bodies start with Euler and Lagrange where
they discussed the restricted problem for $N=3$.
The collinear central configuration was introduced by Euler, whereas triangular central
configuration was introduced by Lagrange. The central problem deals with determination
of the geometric configuration for $N-$point masses interacting in gravitational fields.
Till date an analytical solution for this $N-$body $(N\geq3)$ problem is not available.
Many results have been published and put forward by a number of researchers with various
perturbations like oblateness or triaxial of the primaries(e.g.,\cite{AGM16} \cite{AGV15},
\cite{AAGM15}, \cite{elshaboury})
the radiation pressure effects (e.g., \cite{SAKA18}), effect of the Coriolis and centrifugal
forces (e.g., \cite{BH78}), the restricted three body problem with variable mass
(e.g., \cite{SI83}, \cite{DSI88}, \cite{AM15}) and many others in the context of restricted
three--body problem (e.g., \cite{AGM14}, \cite{AAEA14}, \cite{WHH}).
\bigskip

In spite of all these
facts, the problem involving $N\geq3$ is still interesting and open burning topic of research.
The various surprising results in the study of restricted problem of three bodies paved the
path and motivated researchers to extend this dynamical model into restricted four and
five--body problem. However, as we move on the restricted four and five--body problem from
the restricted three--body problem, the complications and challenges increase manifold.
Some of the notable study in the context of restricted  four--body problem with various
perturbations are (e.g., \cite{SAP17}, \cite{SMA18}), with oblateness of the primaries
(e.g., \cite{SAA18}, \cite{Sur17b}), effect of the Coriolis and centrifugal forces
(e.g., \cite{SV15}, \cite{SAA17},  \cite{agg18}), effect of variable mass
(e.g., \cite{MAB16}, \cite{mit18}).
\bigskip

The restricted problem of five bodies was introduced by \cite{oll88}, where he discussed
the motion of the fifth body of negligible mass, in comparison to remaining four bodies.
His mathematical model was described as follows:  three equal masses primaries moving
around their gravitational center in circular orbit under their mutual gravitational
attraction were taken on the same plane whereas, a mass of $\beta > 0$ times, the mass
of one of the three primary bodies is supposed at the center of mass. The presented
mathematical model of five--body problem reduced to the restricted four--body problem
for particular value of $\beta= 0$. His study unveils the fact that there exist nine
libration points in total in which three are stable for $\beta > 43.18$, on the other
hand all these nine libration points are linearly unstable for smaller values of $\beta$.
\bigskip

In continuation of Oll\"{o}ngren, \cite{PK07} have introduced the effect of radiation
pressure due to some or all of the four primaries and explored numerically that the
number of collinear libration points of this dynamical system depends on mass parameter,
as well as on the radiation pressure.
Most recently, \cite{ZS17} have investigated the basins of convergence associated with
the libration points by using multivariate version of Newton--Raphson iterative scheme
in the restricted five--body problem. The numerical simulation has been presented to
explore the behaviour that how the libration points (which act as attractors) of the
system attract the initial conditions, always referred as nodes lying on the configuration
plane and constitute a domain of basins of convergence.  The author has emphasised that the
geometry of the basins of convergence is highly influenced by the mass parameter.
\bigskip

The aforementioned literatures provide us an idea to introduce the effect of variable mass
in the frame of five--body problem. The effects of variable mass in three or four--body
problem have explored various new results and facts, therefore, the study of the effect
of variable mass within the frame of five--body problem is novel and worth study in spite
of lots of complications.
\bigskip

The manuscript is prepared as follows:  A literature review within frame of $N-$bdy problem
is stated in Section \ref{intro}, but the most important properties and equations of motion
 for five--body problem are discussed in Section \ref{props}. In Section \ref{eqpts},
 the main numerical results
regarding the parametric evolution of the positions of libration points are presented,
while in Section \ref{Stability of libration points} the  stability of these points is studied.
The most intrinsic
properties of the dynamical system of restricted five--body problem have been revealed by using
the Newton-Raphson basins of convergence in Section \ref{basins of convergence}. Finally,
discussion and conclusion are drew in Section \ref{Discussion and conclusions}.
\section{Structures of equations of motion}
\label{props}
 The dynamical system of studying is the circular restricted five--body problem.
This problem consists of four primaries
$P_i$, $i=0, 1, 2, 3$ which move in circular orbit around their common
center of mass. We, further, supposed that the fifth body whose mass is
too small in comparison to masses of the primaries, and its
mass is not constant on the contrary its mass varies with respect to time.
In this context the fifth body (test particle) dose not affect on the
motion of the four primaries.
\begin{figure}
\centering
\includegraphics[scale=0.7]{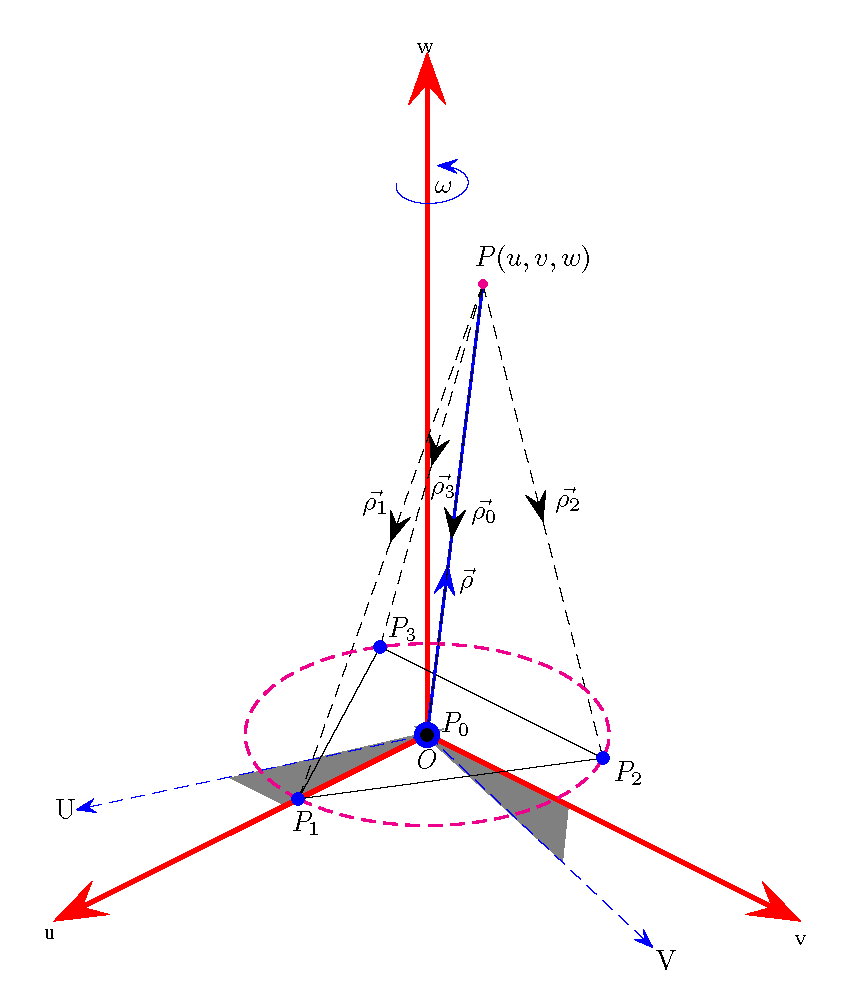}
\caption{\small{The planar configuration of the circular restricted five--body
problem.
The blue dots shows the positions of the four primary bodies.}}
 \label{Fig:1}
\end{figure}

In the planar motion of the test particle, we choose the rotating frame of reference where the origin coincides with the center of mass of the primaries. The positions of the center of the primaries are: $(u_0, v_0)=$$(0, 0)$, $(u_1, v_1, w_1)=$$(1/\sqrt{3}, 0, 0)$, $(u_2, v_2, w_2)=$$(-1/2\sqrt{3}, 1/2, 0)$, and $(u_3, v_3, w_3)=$ $(-1/2\sqrt{3}, -1/2, 0)$, while the dimensionless masses of the primaries are $m_0=\beta m^*$, $m_1=m_2=m_3=m^*=1$. In addition, the three primaries with mass $m^*$ are situated at the vertices of an equilateral triangle whose side is unity, while the fourth primary, with mass $\beta m^*$, is situated at the center of the equilateral triangle.

According to \cite{oll88}, and \cite{ZS17}, in the synodic coordinates system,
the effective potential function of the circular restricted five--body problem is given as:
\begin{equation}\label{Eq:1N}
\Omega^*= k\sum\limits_{i=0}^3 \frac{m_i}{\rho_i} + \frac{1}{2}\left(u^2 + v^2 \right),
\end{equation}
where $k=1/3(1+\beta\sqrt{3})$ and
\begin{align}
\rho_i &=\sqrt{(u-u_i)^2 + (v-v_i)^2 + (w-w_i)^2}, i=0, 1, 2, 3.\nonumber
\end{align}
are the distances between the respective primaries and  test particle.

The equations of motion for a test particle, with dimensionless variables in a
rotating coordinates system in which the primary $m_1$ is fixed on
the \emph{O}u--axis, are read as:
\begin{subequations}
\begin{align}
\label{Eq:4a}
\left(\ddot{u}-2\dot{v}\right)+n\frac{\dot{m}}{m}\left(\dot{u}-v\right)&=\Omega^*_u,\\
\label{Eq:4b}
\left(\ddot{v}+2\dot{u}\right)+n\frac{\dot{m}}{m}\left(\dot{v}+u\right)&=\Omega^*_v,\\
\label{Eq:4c}
\ddot{w}+n\frac{\dot{m}}{m}\dot{w}&=\Omega^*_w,
\end{align}
\end{subequations}
where  $\Omega^*_u,$ $\Omega^*_v$ and $\Omega^*_w$ are partial derivatives
of the effective potential given in Eq.\,\eqref{Eq:1N}.

Moreover, the Jeans' law states that $dm/dt=-\alpha m^s$, where
$\alpha$ is a constant coefficient and $0.4\leq s \leq 4.4$. Acquainting
the space--time transformations which read as:
\[u=\gamma^{-q} x, \,\, v=\gamma^{-q} y, \,\, w=\gamma^{-q} z, \,\, dt=\gamma^{-k}d\tau ,\]
where $\gamma=m/m_{init}$, $m_{init}$ is the mass of the test particle
at the initial time i.e., $t=0$.  Adopting the procedure given by
\cite{SI83} and \cite{DSI88}, to free the equations of motion of
the test particle from the factor which depends upon the variation
of mass, it is sufficient to set $s=1, q=\frac{1}{2}, k=0$.
Therefore, the components of velocity and acceleration can be read as:
\begin{subequations}
\begin{eqnarray}
\label{Eq:5a}
\gamma^{\frac{1}{2}}\dot{u}&=&x^{'}+\frac{1}{2}\alpha x,\\
\label{Eq:5b}
\gamma^{\frac{1}{2}}\dot{v}&=&y^{'}+\frac{1}{2}\alpha y,\\
\label{Eq:5c}
\gamma^{\frac{1}{2}}\dot{w}&=&z^{'}+\frac{1}{2}\alpha z,\\
\label{Eq:5d}
\gamma^{\frac{1}{2}}\ddot{u}&=&x^{''}+\alpha x^{'}+\frac{1}{4}\alpha^2 x,\\
\label{Eq:5e}
\gamma^{\frac{1}{2}}\ddot{v}&=&y^{''}+\alpha y^{'}+\frac{1}{4}\alpha^2 y,\\
\label{Eq:5f}
\gamma^{\frac{1}{2}}\ddot{w}&=&z^{''}+\alpha z^{'}+\frac{1}{4}\alpha^2 z,
\end{eqnarray}
\end{subequations}
where
\begin{eqnarray*}
(')=\frac{d}{d\tau}, \quad (.)=\frac{d}{dt},\quad \text{and}\quad \frac{d}{dt}=\frac{d}{d\tau}.
\end{eqnarray*}
Using Eqs.\,(\ref{Eq:5a}--\ref{Eq:5f}) into Eqs.\,(\ref{Eq:4a}--\ref{Eq:4c}), we get
\begin{subequations}
\begin{align}
\label{Eq:6a}
\left(\ddot{x}-2\dot{y}\right)-\alpha(n-1)\dot{x}+\alpha(n-1)y&=\Omega^{**}_x,\\
\label{Eq:6b}
\left(\ddot{y}+2\dot{x}\right)-\alpha(n-1)\dot{y}-\alpha(n-1)x&=\Omega^{**}_y,\\
\label{Eq:6c}
\ddot{z}-\alpha(n-1)\dot{z}&=\Omega^{**}_z,
\end{align}
\end{subequations}
where
\begin{align*}
\Omega^{**}&= k\gamma^{\frac{3}{2}}\sum\limits_{i=0}^3 \frac{m_i}{r_i}+\frac{\alpha^2}{8}(2n-1)\big(x^2 +y^2+z^2\big)+\frac{1}{2}\Big(x^2+y^2\Big),\nonumber\\
r_i &= \sqrt{(x-x_i)^2 + (y-y_i)^2 + (z-z_i)^2}, \nonumber\\
x_0&=0, x_1=\frac{\gamma^\frac{1}{2}}{\sqrt{3}}= -2x_2=-2x_3,\nonumber\\
y_0&=y_1=0, y_2=\frac{\gamma^\frac{1}{2}}{2}=-y_3,\nonumber\\
z_i&=0, i=0, 1, 2, 3.
\end{align*}

The Eqs.\,(\ref{Eq:6a}--\ref{Eq:6c}) describe the equations of motion of the fifth
body where the variation of mass of the fifth particle is non-isotropic. Further,
it is supposed that the variation of mass is from the entire surface
(i.e., from $n$ distinct points), and the ejaculation from or fall of masses
to the surface has zero momentum in the circular restricted five-body problem.
Furthermore, when we consider the case that the variation of the mass emanate
from one point only (i.e., $n=1$), thus, the equations of motion  given by
Eqs.\,(\ref{Eq:6a}--\ref{Eq:6c}) read as:
\begin{subequations}
\begin{align}
\label{Eq:7a}
\ddot{x}-2\dot{y}&=\Omega_x,\\
\label{Eq:7b}
\ddot{y}+2\dot{x}&=\Omega_y,\\
\label{Eq:7c}
\ddot{z}&=\Omega_z,
\end{align}
\end{subequations}
where
\begin{align}\label{Eq:8}
\Omega&= k\gamma^{\frac{3}{2}}\sum\limits_{i=0}^3 \frac{m_i}{r_i}+\frac{\alpha^2}{8}\left(x^2 +y^2+z^2 \right)+\frac{1}{2}\Big(x^2+y^2\Big),\nonumber\\
\Omega_x&=-k\gamma^\frac{3}{2}\sum_{i=0}^{3}\frac{m_i\tilde{x}_i}{r_i^3}+\Big(1+\frac{\alpha^2}{4}\Big)x,\nonumber\\
\Omega_y&=-k\gamma^\frac{3}{2}\sum_{i=0}^{3}\frac{m_i\tilde{y}_i}{r_i^3}+\Big(1+\frac{\alpha^2}{4}\Big)y,\nonumber\\
\Omega_z&=-k\gamma^\frac{3}{2}\sum_{i=0}^{3}\frac{m_i\tilde{z}_i}{r_i^3}+\frac{\alpha^2}{4}z,\nonumber\\
\tilde{x}_i&=x-x_i,\quad \tilde{y}_i=y-y_i,\quad \tilde{z}_i=z-z_i.\nonumber
\end{align}
In the same vein, the $2^{nd}-$order partial derivatives which will be used
to discuss the linear stability of the obtained libration point can be written as:
\begin{subequations}
\begin{eqnarray}
\label{Eq:7Na}
\Omega_{xx}&=&-k\gamma^\frac{3}{2}\sum_{i=0}^{3}\Big(\frac{m_i}{r_i^3}-\frac{3m_i\tilde{x}_i^2}{r_i^5}\Big)+\Big(1+\frac{\alpha^2}{4}\Big),\\
\label{Eq:7Nb}
\Omega_{yy}&=&-k\gamma^\frac{3}{2}\sum_{i=0}^{3}\Big(\frac{m_i}{r_i^3}-\frac{3m_i\tilde{y}_i^2}{r_i^5}\Big)+\Big(1+\frac{\alpha^2}{4}\Big),\\
\label{Eq:7Nc}
\Omega_{zz}&=&-k\gamma^\frac{3}{2}\sum_{i=0}^{3}\Big(\frac{m_i}{r_i^3}-\frac{3m_i\tilde{z}_i^2}{r_i^3}\Big)+\frac{\alpha^2}{4},\\
\label{Eq:7Nd}
\Omega_{xy}&=&k\gamma^\frac{3}{2}\sum_{i=0}^{3}\frac{3m_i\tilde{x}_i\tilde{y}_i}{r_i^5}=\Omega_{yx},\\
\label{Eq:7Ne}
\Omega_{xz}&=&k\gamma^\frac{3}{2}\sum_{i=0}^{3}\frac{3m_i\tilde{x}_i\tilde{z}_i}{r_i^5}=\Omega_{zx},\\
\label{Eq:7Nf}
\Omega_{yz}&=&k\gamma^\frac{3}{2}\sum_{i=0}^{3}\frac{3m_i\tilde{y}_i\tilde{z}_i}{r_i^5}=\Omega_{zy}.
\end{eqnarray}
\end{subequations}
\section{Equilibrium points}
\label{eqpts}
The equilibrium point exists if and only if the following conditions hold:
\begin{equation}
\dot{x}=\dot{y}=\dot{z}=\ddot{x}=\ddot{y} =\ddot{z} = 0.\nonumber
\end{equation}
Similar to the mass parameter of the classical restricted three--body problem,
we can take a mass parameter $\mu=1/(1+\beta)$ to compare them. Therefore, we
have $\mu\in (0, 1]$ when $\beta \in [0,\infty)$.
\subsection{The libration points in configuration $(x, y)-$plane}
\label{lib:1}
In this subsection, we restrain our analysis only to the equilibrium points
which lie on the $(x, y)-$plane, when $z=0$. The associated positions
$(x_0, y_0)$  of the libration points can easily be found by solving
numerically the system of the $1^{st}-$order partial derivative equations
i.e., Eqs.\,\eqref{Eq:9} appended below:
\begin{equation}
\begin{cases}
\Omega_x(x,y,z)|_{(z=0)}= 0, \\
\Omega_y(x,y,z)|_{(z=0)}= 0. \\
\end{cases}
\label{Eq:9}
\end{equation}
The total number  of the equilibrium points location, in the circular
restricted five--body problem in classical case, depend on the mass parameter
$\mu$ (see \cite{ZS17}). Moreover, the number of the libration points vary
for critical value of the mass parameter(i.e., $\mu^*=0.98617275$).
Therefore, when we have taken the mass of the test particle as variable,
we will explore how the number and positions of the equilibrium points
are effected by the parameters $\alpha$ as well as $\gamma$.
\begin{table}
\begin{tabular}{cccc}
  \hline
  $\mu$ & $\mu^*$ &$\alpha$ & Libration points \\
  \hline
$(0, \mu^*) $ and $[\mu^*, 1) $ &0.95353029&2   & 9 and 15\\
$(0, \mu^*) $ and $[\mu^*, 1) $ &0.97290121&1.5 & 9 and 15\\
$(0, \mu^*) $ and $[\mu^*, 1) $ &0.98510106&0.5& 9 and 15\\
$(0, \mu^*) $ and $[\mu^*, 1) $ &0.98617275&0  & 9 and 15 \\
  \hline
\end{tabular}
\caption{The number of libration points when $\alpha$ varies.}
\label{Table:1}
\end{table}
\begin{figure*}
\centering
\resizebox{\hsize}{!}{\includegraphics{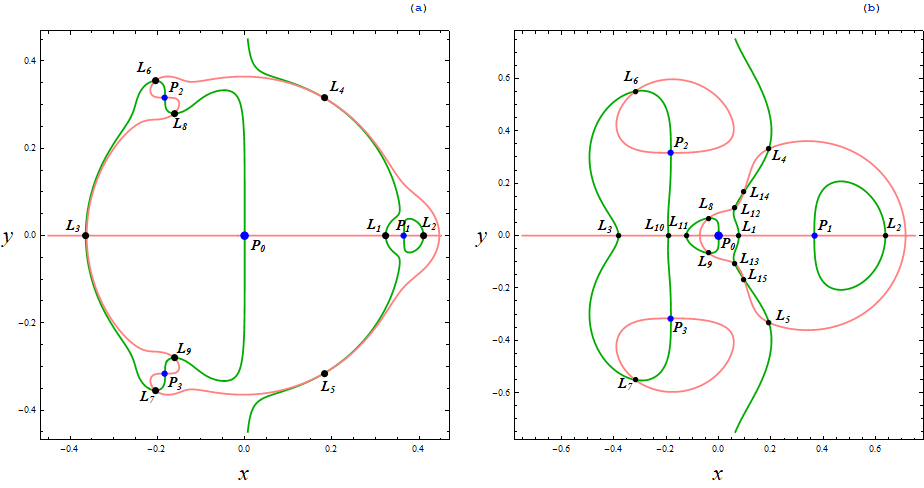}}
\caption{Positions (black dots) and numbering of the equilibrium points ($L_i, i = 1, . . . , 9 \text{or} 15$) through the intersections of $\Omega_x=0$ (green) and $\Omega_y=0$ (blue), when (a--left):$\mu=0.005$, $\alpha=0.2$ and $\gamma=0.4$ (nine equilibrium points), and (b--right): $\mu=0.96353029$, $\alpha=2$ and $\gamma=0.4$ (fifteen equilibrium points). The blue dots denote the centers $(P_i, i = 0, 1, 2, 3)$ of the primaries.}
\label{Fig:2}
\end{figure*}

From  Table\,\eqref{Table:1}, it is revealed that the critical value $\mu^*$
changes, i.e., the interval in which 9 libration points exist decreases and
obviously the length of interval which contain 15 libration points increases
when $\alpha$ increases.

The positions of the equilibrium points can be illustrated by the intersections of
the equations $\Omega_x = 0$, and $\Omega_y = 0$. In Fig.\,\ref{Fig:2}\,(a -- b),
we have shown how the above mentioned equations nail, in every case, the positions
of the equilibrium points, for (a): $\mu= 0.005$ and (b): $\mu = 0.96353029$ with
fixed value of $\alpha=2$ and $\gamma=0.4$. Moreover,  in the corresponding panels
of the figures, we depicted the numbering, $L_i , i = 1, . . . ,9\quad\text{or}\quad 15$,
of all the libration points.
\begin{figure*}
\centering
\resizebox{\hsize}{!}{\includegraphics{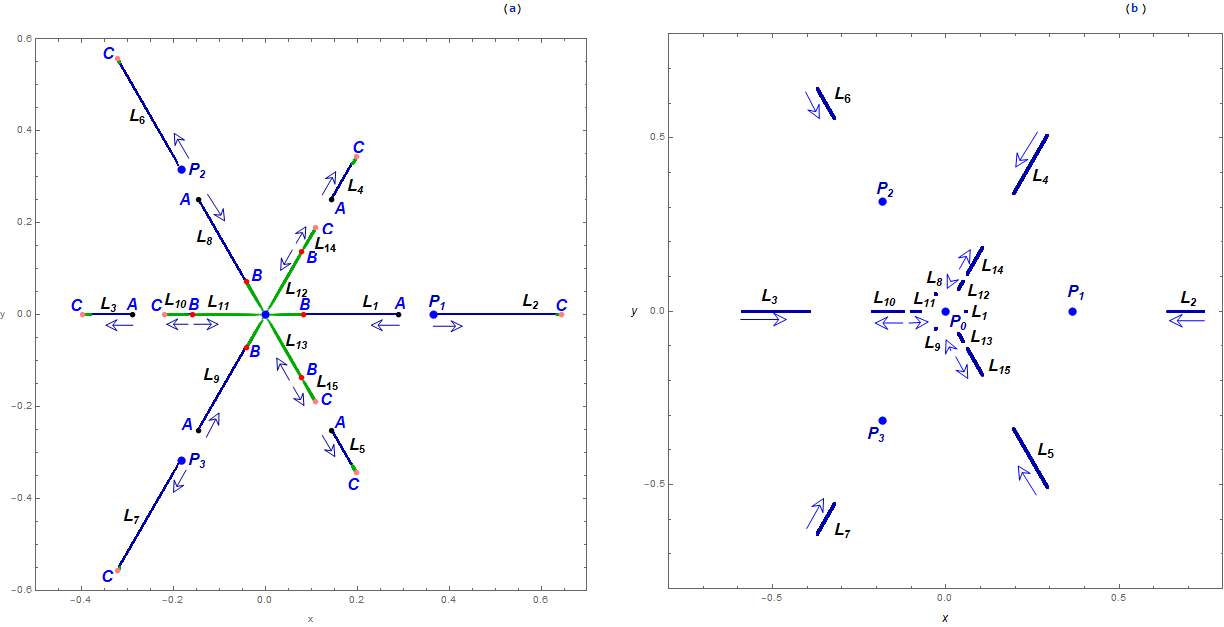}}
\caption{The parametric evolution of the positions of the libration points, $L_i$, $i = 1, . . . , 9 \text{or} 15$, in the restricted five--body problem with variable mass: (a) when $\mu \in (0, 1]$. The arrows indicate the movement direction of the libration points as the value of the mass parameter increases. The big \emph{blue} dots pinpoint the fixed
centers of the primaries, while the small \emph{black, red} and \emph{pink} dots (points $A$, $B$, and $C$)
correspond to $\mu\rightarrow0$, $\mu =\mu^*=0.95353029$, and $\mu= 1$, respectively with $\alpha=2$ and $\gamma=0.4$, (b) when $\mu=0.987$, $\gamma=0.4$, and $\alpha \in(0, 2.2]$. (colour figure online).}
\label{Fig:3}
\end{figure*}

The parametric evolution of the locations of the coplanar [i.e., on $(x, y)-$plane]
equilibrium points are presented in Fig.\,\eqref{Fig:3}, whereas in Fig.\,\eqref{Fig:4}
positions of the out--of--plane [i.e., on the $(x, z)-$plane] libration points are
illustrated. In Fig.\,(\ref{Fig:3}\,a), the movement of the position of libration
points is shown for fixed value of $\alpha=2, \gamma=0.4$ and varying values of
parameter $\mu\in (0, 1]$, whereas in Fig.\,(\ref{Fig:3}\,b),  this movement is
shown for fixed value of $\mu=0.987, \gamma=0.4$ and varying values of $\alpha \in(0, 2.2]$.
From Fig.\,(\ref{Fig:3}\,a), we have observed that when the parameter $\mu$ is just above
zero, the libration points $L_{1, 6, 7}$ emerged in the vicinity of primaries $P_{1,2,3}$
respectively, and the libration points $L_{1, 8, 9, 11, 12 , 13}$ collide with the origin
for $\mu=1$.  If we compare our analysis with Fig.\,(3) of \cite{ZS17}, it is concluded
that the three libration points $L_{1, 8, 9}$ do not emerge in the vicinity of the
primaries when the parameter $\alpha$ due to variable mass is introduced.
It is also noticed that all the libration points germinates with the axes of
symmetry $y=0, y=\sqrt{3\gamma}$ and $y=-\sqrt{3\gamma}$. Moreover, the movement
of the positions of all libration points is same as in Fig.\,(3) of \cite{ZS17}.

In Fig.\,(\ref{Fig:3}\,b), we have observed that the movement of the positions of
the equilibrium points $L_{2, 3, 4, 5, 6, 7}$ is reversed (as these points move
far from the primary $P_0$ along the line of symmetry when $\mu$ increases, see
Fig.\,(\ref{Fig:3}\,a) and it started to move toward the primary $P_0$ along the
line of symmetry when $\alpha$ increases. In addition, the change in the libration
points $L_{1, 8, 9}$ is negligible whereas $L_{10, 14, 15}$ move away from primary
$P_0$ as $\alpha$ increases.
\subsection{Out--of--plane libration points}
\label{Out-of-plane libration points}
\begin{figure}
\centering
\resizebox{\hsize}{!}{\includegraphics{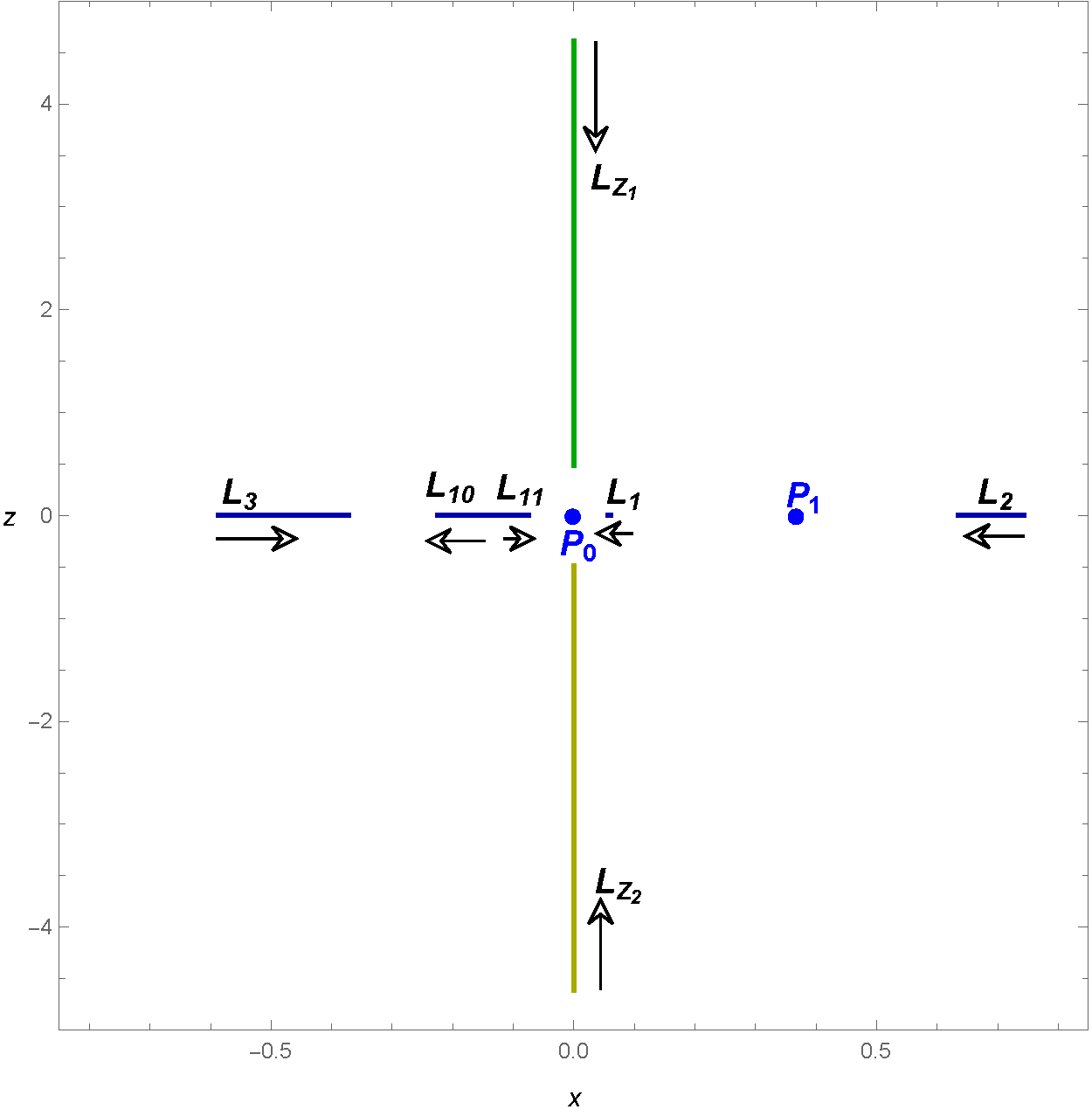}}
\caption{The parametric evolution of the positions of the out--of--plane libration points, $L_{z_1}$, and $L_{z_2}$,  in the restricted five--body problem with variable mass when $\mu= 0.9862727, \gamma= 0.4$ and $\alpha \in(0, 2.2]$. The arrows indicate the movement direction of the libration points as the value of the mass parameter increases. The big \emph{blue} dots pinpoint the fixed centers of the primaries. (colour figure online).}
\label{Fig:4}
\end{figure}

In this subsection, we continue our analysis with the out--of--plane equilibrium
points, i.e., the libration points which lie on $(x, z)-$plane ($y=0$). By solving
numerically the system of $1^{st}-$order derivative equations, with a help of
Eqs.\,\eqref{Eq:10}, we obtain
\begin{equation}
\begin{cases}
\Omega_x(x,y,z)|_{(y=0)}=0, \\
\Omega_z(x,y,z)|_{(y=0)}=0,
\end{cases}
\label{Eq:10}
\end{equation}
we can determine the locations of the out--of--plane equilibrium points.
The intersections of the equations $\Omega_x=0$, and $\Omega_z=0$ describe
the locations of these equilibrium point. In Fig.\,\eqref{Fig:4}, the parametric
evolution of the locations of the libration points on $(x, z)-$plane, when
$\alpha \in (0, 2.2]$, is illustrated for pre defined value of the parameter
$\mu= 0.9862727$, $\gamma= 0.4$ and varying value of $\alpha \in (0,2.2]$.
As the value of the parameter $\alpha>0$ increases, a pair of symmetrical
(with respect to $x-$axis) out--of--plane libration point namely $L_{z_1}$
and $L_{z_2}$ appear on the $z-$axis. In addition, these equilibrium points
move towards the central primary $P_0$ as the parameter $\alpha$ increases.
Finally, it is unveiled that the libration points always lie on coordinates
axes $(x, z)$.
\section{Stability of libration points}
\label{Stability of libration points}

The dynamical systems which describe the restricted five--body problem
are developed, but they do not provide a concise characterization relating to
the fifth body motion. The measurements process and the behaviour of
dynamical motion of these systems are affected by the parameters variation
or the state variables which give an exact definitions of the initial conditions.
In addition, there is an extra difficulty to find a solution for these systems
directly, for any parameter selection from a specific measurements set.
Regard to the large complexity that included in these systems, our attentions
are paid to linearize the dynamical system in Eqs.\,(\ref{Eq:7a} -- \ref{Eq:7c})
to obtain more simple dynamical system that can be used to underline the features
of fifth body motion and its dynamical characterizations.
To understand and investigate the dynamics of possible motion of the fifth body
in the proximity of libration points, the equations of motion, we have linearized
Eqs.\,(\ref{Eq:7a} -- \ref{Eq:7c}) along the initial state vector.
Thereby, we expand their right hand-side  around the equilibria points.
Hence, the obtained linear system  is called the \emph{variational equations}.
Applying the procedure of \cite{MAB16, mit18}, we shall give displacements
in $(x_0, y_0, z_0)$ as:
\begin{eqnarray}\nonumber
x=x_0+\epsilon_1, \quad
y=y_0+\epsilon_2, \quad
z=z_0+\epsilon_3, (\epsilon_1, \epsilon_2, \epsilon_3 <<1)
\end{eqnarray}
where $(x_0, y_0, z_0)$  denote the position of equilibrium point
for a fixed value of time $t$. The associated variational equations
can be written as:
\begin{eqnarray}\label{eqn:13}
\ddot{\epsilon_1}-2\dot{\epsilon_2} &=& (\Omega_{xx})_0\epsilon_1 +(\Omega_{xy})_0\epsilon_2 +(\Omega_{xz})_0\epsilon_3 \nonumber, \\
\ddot{\epsilon_2}+2\dot{\epsilon_1} &=& (\Omega_{yx})_0\epsilon_1 +(\Omega_{yy})_0\epsilon_2 +(\Omega_{yz})_0\epsilon_3 \nonumber, \\
\ddot{\epsilon_3} &=& (\Omega_{zx})_0\epsilon_1 +(\Omega_{zy})_0\epsilon_2 +(\Omega_{zz})_0\epsilon_3,
\end{eqnarray}
where the subscript `$0$' in Eqs.\,\ref{eqn:13} associated with the values
of $2^{nd}-$order partial derivatives of $\Omega$ evaluated at the libration
point $(x_0, y_0, z_0)$ under consideration. The problem of constant mass
can be easily obtained by taking $\alpha=0$ in the problem of variable mass.

Applying the procedure and transformations given in \cite{MAB16, mit18},
the characteristic equation of the coefficient matrix is written as
\begin{eqnarray}\label{Eq:11N}
 \lambda^{6}&-&3\alpha \lambda^{5}+\Big(\frac{15}{4}\alpha^{2}+\phi_1\Big)\lambda^{4}-\Big(\frac{5}{2}\alpha^{3}+2\phi_1\alpha\Big)\lambda^{3}\nonumber\\
 &&+\Big(\frac{15}{16}\alpha^{4}+\frac{3}{2}\phi_1\alpha^{2}+\phi_2\Big)\lambda^{2}-\Big(\frac{3}{16}\alpha^{5}+\frac{1}{2}\phi_1\alpha^{3}\nonumber\\
 &&+\phi_2\alpha\Big)\lambda+\Big(\frac{1}{64}\alpha^{6}+\frac{1}{16}\phi_1\alpha^{4}+\frac{1}{4}\phi_2\alpha^{2}+\phi_3\Big)=0,
\end{eqnarray}
where
\begin{eqnarray}
  \phi_1 &=& 4-(\Omega_{xx})_0-(\Omega_{yy})_0-(\Omega_{zz})_0,\nonumber\\
  \phi_2 &=& (\Omega_{xx})_0(\Omega_{zz})_0+(\Omega_{yy})_0(\Omega_{zz})_0+(\Omega_{xx})_0 (\Omega_{yy})_0-4(\Omega_{zz})_0\nonumber\\
  &&-[(\Omega_{xy})_0]^{2}-[(\Omega_{xz})_0]^{2}-[(\Omega_{yz})_0]^{2},\nonumber\\
  \phi_3 &=& -(\Omega_{xx})_0(\Omega_{yy})_0(\Omega_{zz})_0+(\Omega_{zz})_0[(\Omega_{xy})_0]^{2}
\nonumber\\
  &&+(\Omega_{yy})_0[(\Omega_{xz})_0]^{2}
  +(\Omega_{xx})_0[(\Omega_{yz})_0]^{2}-2(\Omega_{xy})_0\times(\Omega_{xz})_0(\Omega_{yz})_0,\nonumber
\end{eqnarray}
the values of $(\Omega_{xx})_0$, $(\Omega_{yy})_0$, $(\Omega_{zz})_0$,
$(\Omega_{xy})_0$, $(\Omega_{xz})_0$ and $(\Omega_{yz})_0$ are given by the
Eqs.\,(\ref{Eq:7Na} -- \ref{Eq:7Nf}).

If the exact positions of the in plane [i.e., on $(x, y)-$plane] and out--of--plane
[i.e., on $(x, z)-$plane] libration points are denoted by $(x_0, y_0, 0)$ and
$(x_0, 0, z_0)$, respectively, then we can easily decide the linear stability of
these libration points by determining the nature of the roots of the characteristic
equation [i.e., Eq.\,\eqref{Eq:11N}]. We have numerically determined the linear
stability of the libration points for various combination of the parameters,
and found each libration point is unstable.
\section{Basins of convergence}
\label{basins of convergence}
\begin{figure*}
\centering
\resizebox{\hsize}{!}{\includegraphics{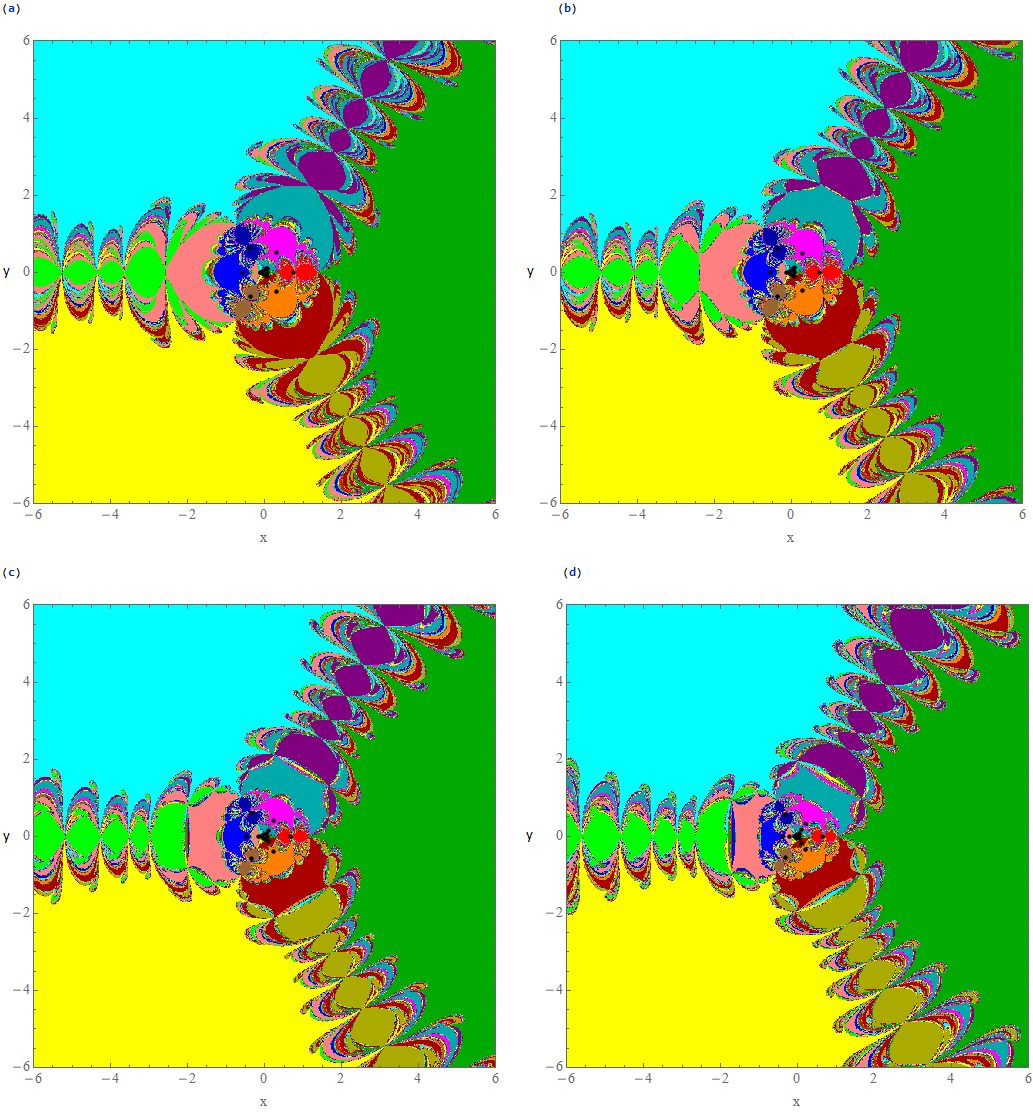}}
\caption{The basins of attraction for fixed value of $\gamma=0.4$ and $\mu=0.986173$. (a) $\alpha=0.2$; (b) $\alpha=0.75$; (c) $\alpha=1.5$; (d) $\alpha=2$.  The color code for the libration points $L_1$,...,$L_{15}$ is as follows: $L_{1}\emph{(green)}$; $L_{2}\emph{(red)}$; $L_{3}\emph{(blue)}$; $L_{4}\emph{(magenta)}$; $L_{5}\emph{(orange)}$; $L_{6}\emph{(indigo)}$; $L_{7}\emph{(brown)}$; $L_{8}\emph{(cyan)}$; $L_{9}\emph{(yellow)}$; $L_{10}\emph{(pink)}$; $L_{11}\emph{(fluorescent green)}$;  $L_{12}\emph{(purple)}$;   $L_{13}\emph{(olive)}$;   $L_{14}\emph{(teal)}$;  $L_{15}\emph{(crimson)}$;    and non--converging points (white). (colour figure online).}
\label{Fig:5}
\end{figure*}
\begin{figure*}
\centering
\resizebox{\hsize}{!}{\includegraphics{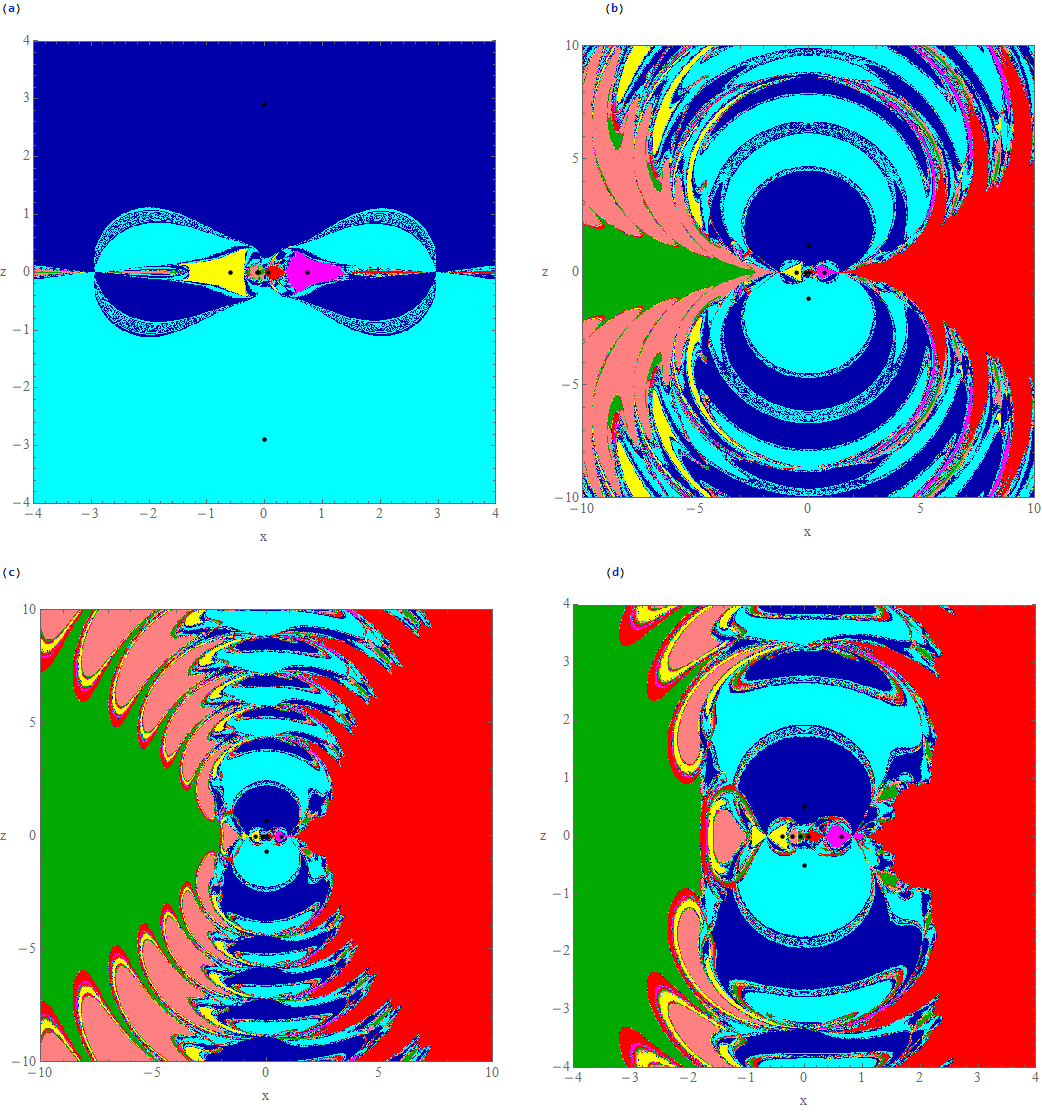}}
\caption{The out-of-plane basins of attraction for fixed value of $\gamma=0.4$ and $\mu=0.9862727$. (a) $\alpha=0.2$; (b) $\alpha=0.75$; (c) $\alpha=1.5$; (d) $\alpha=2$. The color code for the libration points is as follows: $L_{1}\emph{(red)}$; $L_{2}\emph{(magenta)}$; $L_{3}\emph{(yellow)}$; $L_{10}\emph{(pink)}$; $L_{11}\emph{(fluorescent green)}$;  $L_{z_1}\emph{(blue)}$;   $L_{z_2}\emph{(cyan)}$; and non-converging points (white). (colour figure online).}
\label{Fig:6}
\end{figure*}
\begin{figure*}
\centering
\resizebox{\hsize}{!}{\includegraphics{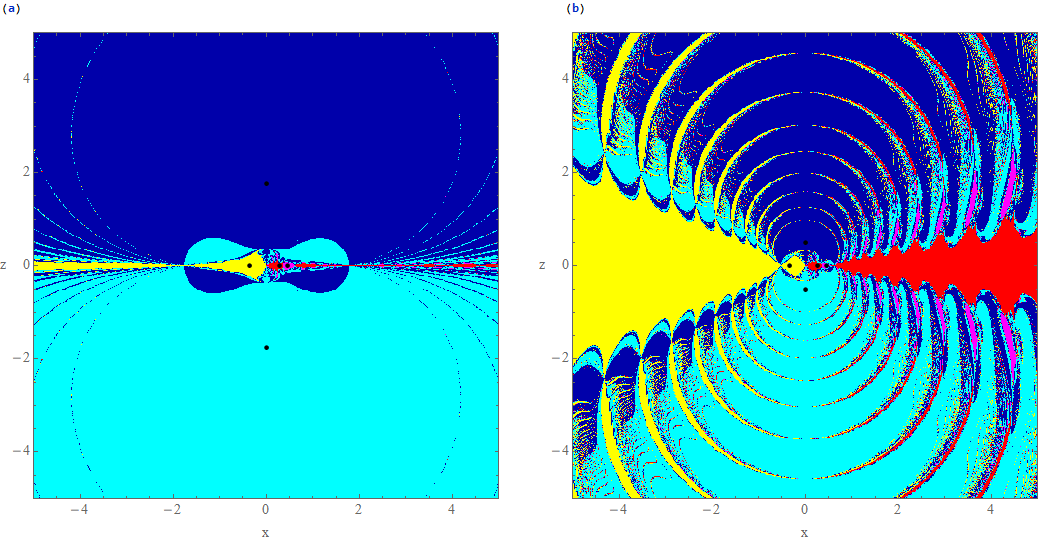}}
\caption{The out-of-plane basins of attraction for fixed value of
$\gamma=0.4$ and $\mu=0.05$. (a) $\alpha=0.2$; (b) $\alpha=1.25$. The color code for the libration points is as follows: $L_{1}\emph{(red)}$; $L_{2}\emph{(magenta)}$; $L_{3}\emph{(yellow)}$; $L_{z_1}\emph{(blue)}$;   $L_{z_2}\emph{(cyan)}$; and non-converging points (white). (colour figure online).}
\label{Fig:7}
\end{figure*}
In spite of various available iterative scheme to solve the system of non--linear
equations, the Newton--Raphson iterative scheme has considered as one of the most
enthralled as well as precise iterative method to solve these equations.
We can solve the system of multivariate functions $F(\mathbf{x}) = 0$ by applying
the multivariate iterative scheme appended below:
\begin{equation}\label{Eq:401}
 \textbf{x}_{n+1}=\textbf{x}_n-{J}^{-1}f(\textbf{x}_n),
\end{equation}
where $f(\textbf{x}_n)$ represents the system of equations, while $J^{-1}$
represents the corresponding inverse Jacobian matrix [see Eq.\,\eqref{Eq:401}].
In the recent time, the study of the basins of convergence by using the multivariate
version of the Newton--Raphson iterative scheme are present in various dynamical
system (e.g., \cite{SMA18}, \cite{SZK18}, \cite{Z18}, \cite{ZSMA18}).
In our system, we have three equations, i.e., $\Omega_x=\Omega_y=\Omega_z=0$. It can be
noticed that the Newton--Raphson iterative scheme is applicable in system of three
equations but it is very complicated. Therefore, to make the iterative scheme simple,
we have bifurcated our study into two part: the libration points on $(x, y)-$ plane and
the out--of--plane libration points which lie on $(x, z)-$ plane.
Thus, the bivariate Newton--Raphson iterative scheme can be used on the system:
\begin{subequations}
\begin{eqnarray}
 \Omega_x(x, y, 0)&=&0, \nonumber\\
 \Omega_y(x, y, 0)&=&0.\nonumber
\end{eqnarray}
\end{subequations}
Moreover, the iterative formulae for the $(x, y)$ plane can be written as:
\begin{subequations}
\begin{eqnarray}
\label{Eq:403a}
x_{n+1}&=&x_n-\frac{\Omega_{x_n}\Omega_{y_ny_n}-\Omega_{y_n}\Omega_{x_ny_n}}{\Omega_{x_nx_n}\Omega_{y_ny_n}-\Omega_{x_ny_n}\Omega_{y_nx_n}},\\
\label{Eq:403b}
y_{n+1}&=&y_n+\frac{\Omega_{x_n}\Omega_{y_nx_n}-\Omega_{y_n}\Omega_{x_nx_n}}{\Omega_{x_nx_n}\Omega_{y_ny_n}-\Omega_{x_ny_n}\Omega_{y_nx_n}}.
\end{eqnarray}
\end{subequations}
In the same vein, the bivariate Newton-Raphson iterative scheme can be used on the system:
\begin{eqnarray}
 \Omega_x(x, 0, z)&=&0, \nonumber\\
 \Omega_z(x, 0, z)&=&0.\nonumber
\end{eqnarray}
Therefore, the iterative formulae for the $(x, z)$ plane is read as:
\begin{subequations}
\begin{eqnarray}
\label{Eq:404a}
x_{n+1}&=&x_n-\frac{\Omega_{x_n}\Omega_{z_nz_n}-\Omega_{z_n}\Omega_{x_nz_n}}{\Omega_{x_nx_n}\Omega_{z_nz_n}-\Omega_{x_nz_n}\Omega_{z_nx_n}},\\
\label{Eq:404b}
z_{n+1}&=&z_n+\frac{\Omega_{x_n}\Omega_{z_nx_n}-\Omega_{z_n}\Omega_{x_nx_n}}{\Omega_{x_nx_n}\Omega_{z_nz_n}-\Omega_{x_nz_n}\Omega_{z_nx_n}}.
\end{eqnarray}
\end{subequations}
where the values of $x, y$ and $z$ coordinates at the $n-$th step of the iterative
scheme are $x_n, y_n$ and $z_n$ respectively, in the Newton--Raphson scheme, See
Eqs.\,(\ref{Eq:403a}, \ref{Eq:403b}, \ref{Eq:404a} and \ref{Eq:404b}). Moreover,
the corresponding partial derivatives of the potential function are represented
by the subscripts of $\Omega(x, y, z)$.

In this subsections, we discuss how the parameter $\alpha$ affects the topology of
the domain of the basins of convergence in the restricted problem of five bodies
with variable mass by taking two cases with respect to the type of plane.
The color coded diagrams are used to classify the nodes on the different
type of plane where each pixel is associated with unlike color, corresponding
to the final attractor of the linked initial conditions.

The used iterative scheme, ie., Newton--Raphson method, works under the following
philosophy: the initial conditions $(x_0, y_0)$ or $(x_0, z_0)$  activates the
iterative scheme, which ends when the iterative procedure reached to one of the
equilibrium point (attractor) with predefined accuracy. We assume that the numerical
method converges for a particular initial condition if it results to one of the
equilibrium points of the system for that particular initial condition.
The collection of all the initial conditions, which converge to same
attractors, compile the basins of convergence or attracting regions.

To reveal the topology of the basins of convergence, a double scan of the
$(x, y)$ and $(x, z)-$planes is performed. Moreover, in each plane, we specify
a dense grid of $1024\times 1024$ nodes to be used as an initial conditions of
the Newton--Raphson iterative method. The maximum number of iterations for the
iterative scheme is set to $N_{max}=500$, whereas, the iterative scheme stop
only when an attractor is reached, with predefined accuracy of $10^{-15}$.
\subsection{Results for the $(x, y)-$ plane}
\label{Results for the $(x, y)$ plane}
In this case, where $\mu=0.986173$, there exist fifteen equilibrium points
in which five are collinear and ten are non-collinear. The domain of the basins
of convergence for the four values of parameter $\alpha$ are depicted in
Fig.\,\eqref{Fig:5}. It is unveiled that the domain of the basins of convergence,
linked with the fifteen equilibrium points, have infinite extent, which together
resemble with the shape of butterfly wings.  Moreover, it is observed that the
whole pattern, i.e., the overall geometry of the configuration plane compiled
of different basins of convergence shrinks rapidly as the value of parameter
$\alpha$ increases. Moreover, the neighbourhood of the basins boundaries are
highly chaotic which are composed of mixtures of initial conditions. It is
unveiled that the topology of the basins of convergence is not very sensitive
with the change in the parameter $\alpha$, however these basins boundaries
changes rapidly with the change in the mass parameter  $\mu$ (see,\cite{ZS17}).

The some of the notable change can be summarized as follows:
\begin{itemize}
  \item [-] The domain of the basins of convergence associated with
  the equilibrium points $L_{2, 6, 7}$ look like the exotic bugs with
  many legs and antennas which exists in the interior region.
  \item [-] Three butterfly wings shaped region originates in the
  neighbourhood of the boundary of the interior regions whose extent
  is infinite. These three butterfly wings are composed of the initial
  conditions in which each wings is mostly occupied by those initial
  condition which converges to  $L_{10, 11}$, $L_{12, 14}$ and
  $L_{13, 15}$, respectively.
  \item [-] We observed that the boundary of the interior region is highly
  chaotic which is composed of the initial conditions, therefore it is inconceivable
  to anticipate which initial condition will converge to which of the attractors.
  \item [-] The whole $(x, y)-$plane is occupied by the initial condition which
  converges to $L_1$, $L_8$, and $L_9$ (see, \emph{green, cyan} and \emph{yellow} regions)
  except the interior region and three butterfly wings. Moreover, these
  three regions are symmetrical with respect to $x-$axis.
  \item [-] It is observed that as we increase the value of the parameter
  $\alpha$, the anterior wing i.e., near the boundary of the interior region
  becomes flatter.
\end{itemize}
\subsection{Results for the $(x, z)-$ plane}
\label{Results for the $(x, z)$ plane}
In this subsection, we discuss the results obtained by numerical simulation
with the $(x, z)-$plane where all the out--of--plane libration points lie.
The topology of the basins of convergence linked with the out--of--plane
equilibrium points is illustrated in Figs.\,(\ref{Fig:6},\,\ref{Fig:7}).
We may observe that the $(x, z)-$plane is covered by several well formed
basins of convergence with infinite extinct.
In Fig.\,\eqref{Fig:6} ( for $\mu$=0.9862727), the basins of convergence
are plotted for four specific increasing values of parameter $\alpha$.
The most notable changes which are associated with the $(x, z)-$plane
for the increasing values of $\alpha$ can be summarized as follows:
\begin{itemize}
  \item The area of the domain of  basins of convergence, linked with the
  collinear libration points $L_{2, 3}$ decreases and $L_{1, 10, 11}$ increases
  rapidly, while the area of the domain of basins of convergence associated with
  the out--of--plane equilibrium points $L_{z_1, z_2}$ decreases rapidly.
  \item The shape of the domain of the basins of convergence linked with the
  equilibrium points changes drastically when the  value of parameter $\alpha$ increases.
  \item The domain of basins of convergence linked with the out--of--plane
  libration points are symmetrical with respect to $x-$axis.
  \item The domain of the basins of convergence connected to equilibrium
  points $L_{2, 3}$ converted into exotic bugs shaped region with many
  legs and antenna for the extremely large value of $\alpha$.
\end{itemize}
In Fig.\,\eqref{Fig:7} (for $\mu=0.05$), the basins of convergence are
illustrated for two increasing values of parameter $\alpha$. We can observe
that the geometry of the basins of convergence alters drastically with the
increase in parameter $\alpha$. For this value of mass parameter $\mu$ there
exist only three collinear equilibrium points, moreover, in this case the
extent of the domain of basins of convergence linked with equilibrium points
are also infinite.  We may observe that as the value of the parameter
$\alpha$ increases, the domain of the basins of convergence connected
with the out--of--plane equilibrium points decreases rapidly and now
(see Fig.\,(\ref{Fig:7}\,b) these domain of the basins of convergence
looks like butterfly wings. Moreover, these butterfly wings shaped regions
are separated by a thin strip which is composed of highly chaotic mixtures
of initial conditions.
As we increase the value of the parameter $\alpha$, it is observed that the
domain of the basins of convergence connected with the equilibrium points
$L_1$ (\emph{red} color) and $L_2$ (\emph{yellow} color) increase rapidly
and hence the domain of basins of convergence linked with the out--of--plane
equilibrium points $L_{z_1}$ and $L_{z_2}$ decreases.
\section{Discussion and conclusions}
\label{Discussion and conclusions}
The existence and stability of the equilibrium points in the circular
restricted five--body problem are studied numerically, when the mass variation of
the fifth body is non--isotropic. In this context the domain of basins
of convergence connected with these points, in-plane and out-of plane
is studied and investigated too. Specifically, we have also numerically
explored that how the parameters $\alpha$ and $\mu$ influences the positions
and the linear stability of the libration points.

The multivariate version of the Newton-Raphson iterative method
is used to discuss the influence of parameter $\alpha$ on the geometry of the
domain of basins of convergence on the configuration $(x, y)-$plane and
$(x, z)-$plane. We may argue that these attracting domain provides various
information as they describe how the points on the configuration $(x,y)-$plane
and $(x, z)-$plane are attracted by attractors which are the libration points
of the dynamical system.  We successfully managed to supervise how the domain
of convergence evolves as the function of the parameter $\alpha$.

In addition the important results can be summarized as follows:
\begin{itemize}
\item The existence and the total number of the libration points depends
strongly on the parameter $\alpha$.
\item The length of the interval which contains nine libration points decreases
while the length of interval which contain fifteen libration points increases
with the increase in the value of parameter$\alpha$.
\item The critical value of mass ratio $\mu^*$ is function of parameter $\alpha$
\item For the value of the parameter $\alpha > 0$,  a pair of symmetrical
 (with respect to $x-$axis) out--of--plane libration points exist on the
 $z-$axis which move towards the primary $P_0$ as the parameter $\alpha$ increases.
\item The stability analysis revealed that none of the libration points in either
  $(x, y)-$plane or in $(x, z)-$plane are linearly stable when the mass of the test
  particle is variable while some of the libration points i.e., $L_{3, 4, 5}$ were
  stable in classical circular five--body problem (see, \cite{ZS17}) for the very
  small values of mass parameter.
\item The domain of convergence corresponding to the libration points in the
  configuration $(x,y)-$plane, extend to infinity, in all the studied values
  of the parameters. In addition, the convergence diagrams of the studied system
  maintained symmetry on the $(x, y)-$plane along the line $x=2\pi/3$ .
\item The attracting domains, associated to out--of--plane equilibrium points
  also extend to infinity, in all the mentioned cases. In this case, the domain
  of convergence on the $(x, z)$ plane is symmetrical about the $x-$axis.
\item The categorisation of the nodes on the $(x,y)-$ and $(x, z)-$ planes
  revealed that none of the points are non--converging in nature, however for
  the very close value of mass parameter $\mu$ to the critical value $\mu^*$,
  it is observed that some of these nodes are very slow converging initial conditions.
\end{itemize}

Finally, we would like to refer to the whole numerical calculation and
the associated graphical illustration are constructed by \emph{the codes of Mathematica
software}. We may argue that the presented numerical analysis and discussed results
may be very useful in the context of the basins of convergence in dynamical systems.
It is worth studying the similarities and the differences, associated with the domains
of the basin of convergence in the five--body problem with variable mass by applying
various other iterative schemes other than the Newton-Raphson iterative method.
\section*{Acknowledgments}
\footnotesize
\begin{description}
  \item[*] The authors are thankful to Center for Fundamental Research in Space dynamics and Celestial mechanics (CFRSC), New Delhi, India for providing research facilities.
\item[*] The authors would like to express their warmest thanks and regards to the anonymous referee for the careful reading of the manuscript and for all the apt suggestions and comments which allowed us to improve both the quality and the clarity of the paper.
\end{description}
\par
\textbf{Compliance with Ethical Standards}
\begin{description}
  \item[-] Funding: The authors state that they have not received any research grants.
  \item[-] Conflict of interest: The authors declare that they have no conflict of interest.
\end{description}


\begin{thebibliography}{}
\bibitem[Aggarwal et al.  (2018)]{agg18}
Aggarwal, R.,  Mittal, A.,   Suraj, M. S.,   Bisht, V., The effect of small
perturbations in the Coriolis and centrifugal forces on the existence of
libration points in the restricted fourâ€body problem with variable mass.
\emph{Astronomical notes}, \textbf{339}(6), 492-512 (2018). doi.org/10.1002/asna.201813411.
%
\bibitem[Abouelmagd and Mostafa (2015)]{AM15}Abouelmagd, E.I., Mostafa, A.
Out of plane equilibrium points locations and the forbidden movement regions
in the restricted three-body problem with variable mass.
\emph{Astrophys. Space Sci.}, \textbf{357} (1), 58 (2015)
%
\bibitem[Abouelmagd et al (2015)]{AAGM15}Abouelmagd, E.I., Alhothuali, M.S.,
Guirao, J.L.G., Malaikah, H.M.: The effect of zonal harmonic coefficients
in the framework of the restricted three-body problem.
\emph{Advances in Space Research}, \textbf{55 }(6), 1660--1672 (2015).
%
\bibitem[Abouelmagd et al (2015b)]{AGV15}Abouelmagd, E.I., Guirao, J.L.G., Vera, J.A.:
Dynamics of a dumbbell satellite under the zonal harmonic effect of an oblate body.
\emph{Communications in Nonlinear Science and Numerical Simulation}, \textbf{20}(3),  1057--1069 (2015).
%
\bibitem[Abouelmagd et al (2014a)]{AAEA14} Abouelmagd E.I., Awad, M.E., Elzayat, E.M.A., Abbas, I.A.:
Reduction the secular solution to periodic solution in the generalized restricted three-body problem.
\emph{Astrophys Space Sci.}, \textbf{350} (2), 495--505 (2014a).
%
\bibitem[Abouelmagd et al (2014b)]{AGM14}Abouelmagd, E.I., Guirao, J.L.G., Mostafa, A.:
Numerical integration of the restricted three-body problem with Lie series,
\emph{Astrophys. Space Sci.}, \textbf{354 }(2),  369--378 (2014b).
%
\bibitem[Abouelmagd and Guirao(2016)]{AGM16} Abouelmagd, E. I. Guirao J. L. G.,
On the perturbed restricted three-body problem, \emph{Applied Mathematics and Nonlinear Sciences}, \textbf{1}(1),  123--144 (2016).
%
\bibitem{BH78}Bhatnagar, K.B., Hallan, P.P. Effect of perturbations in Coriolis and
centrifugal forces on the stability of libration points in the restricted problem,
\emph{Celestial Mechanics}, \textbf{18}, 105  (1978). https://doi.org/10.1007/BF01228710
%
\bibitem[Das et al. (1988)]{DSI88} Dass, R.K., Shrivastava, A.K., Ishwar, B.:
Equations of motion of elliptic restricted problem of three bodies with variable mass.
\emph{Celest. Mech. and Dyn. Astron.,} \textbf{45}(4), 387-393 (1988).
%
\bibitem{esgl}Eduardo, S. G. L., On the central configurations of the planar restricted four--body problem.
       \emph{J. Differential Equations,} \textbf{226}, 323 -- 351 (2006).
%
\bibitem[Elshaboury et al. (2016)]{elshaboury} Elshaboury, S. M., Abouelmagd, E. I., Kalantonis, V. S., Perdios, E. A.,
        The planar restricted three-body problem when both primaries are triaxial rigid bodies:
        Equilibrium points and periodic orbits. \emph{Astrophys. Space Sci.} \textbf{361 }(9), 315 (2016)
%
\bibitem{jaume}Llibre J.,  Mello L. F., New central configurations for the planar 7--body problem.
         \emph{Nonlinear Analysis: Real World Applications.} \textbf{10 }, 2246 -- 2255 (2009).
\bibitem{MAB16} Mittal, A.,
    Aggarwal, R., Suraj, M.S., Bisht, V.S.: Stability of libration points in the restricted four-body problem with variable mass, \emph{Astrophys. Space Sci.}, \textbf{361}, 329 (2016).
\bibitem[Mittal et al. (2018)]{mit18}Mittal, A., Aggarwal, R., Suraj, M.S.,  Arora, M., On the photo-gravitational restricted four-body problem with variable mass, \emph{Astrophys. Space Sci.,} \textbf{363}, 109 (2018).
\bibitem{PK07} Papadakis, K.E., Kanavos, S.S.: Numerical exploration of the photogravitational
restricted five-body problem. \emph{Astrophys. Space
Sci.}, \textbf{310}, 119--130 (2007)
\bibitem{oll88} Oll\"{o}ngren, A., On a particular restricted five-body problem, an analysis
with computer algebra.\emph{ J. Symb. Comput.}, \textbf{6 }, 117--126 (1988)
\bibitem[Shrivastava and Ishwar (1983)]{SI83} Shrivastava, A.K., Ishwar, B.: Equations of motion of the restricted problem of three bodies with variable mass. \emph{Celest. Mech. and Dyn. Astron.,} \textbf{30}, 323-328  (1983)
\bibitem{SV15} Singh, J., Vincent, A.E.: Effect of perturbations in the Coriolis and centrifugal forces on the stability of equilibrium points in the restricted four-body problem. \emph{Few-Body Syst.,} \textbf{56}, 713--723 (2015). DOI 10.1007/s00601-015-1019-3
\bibitem{SAP17} Suraj, M.S., Asique, M.C., Prasad, U., Hassan, M.R., Shalini, K.: Fractal basins of attraction in the restricted four-body problem when the primaries are triaxial rigid bodies. \emph{Astrophys. Space Sci.}, \textbf{362}, 211 (2017)
\bibitem{SAA17} Suraj, M.S.,
    Aggarwal, R.,   Arora, M.: On the restricted four-body problem  with the effect of small perturbations in the Coriolis and centrifugal forces. \emph{Astrophys. Space Sci.}, \textbf{362}, 159 (2017).
\bibitem[\protect\citeauthoryear{Suraj et al.}{2017b}]{Sur17b} Suraj, M.S., Asique, M.C., Prasad, U. et al.: Fractal basins of attraction in the restricted four-body problem when the primaries are triaxial rigid bodies.  \emph{Astrophys Space Sci., }\textbf{362}, 211 (2017b).
\bibitem[\protect\citeauthoryear{Suraj et al.}{2018a}]{SMA18}Suraj, M.S., Zotos, E.E.,  Aggarwal, R., Mittal, A.:
Unveiling the basins of convergence in the pseudo-Newtonian planar circular restricted four-body problem, \emph{New Astronomy,} \textbf{66}, 52--67,  (2018a)
\bibitem{SZK18} Suraj, M.S., Zotos, E.E., Kaur, C., Aggarwal, R., et al.: Fractal basins of convergence of libration points in the planar Copenhagen problem with a repulsive quasi-homogeneous Manev--type potential.  \emph{Int. J. Non-Linear Mech.}, \textbf{103}, 113-127, (2018b)
%
\bibitem{SAA18} Suraj, M.S., Mittal, A., Arora, M. et al.: Exploring the fractal basins of convergence in the restricted four-body problem with oblateness.  \emph{Int. J. Non-Linear Mech.}, \textbf{102}, 62--71 (2018c)
\bibitem[Suraj et al (2018d)]{SAKA18}Suraj, M.S., Aggarwal, R., Kumari, S., Asique, M.C.: Out-of-plane equilibrium points and regions of motion in the photogravitational R3BP when the primaries are hetrogeneous spheroid with three layers. \emph{New Astronomy}, \textbf{63}, 15--26 (2018d).
\bibitem{WHH}Wallin, J.F., Holincheck, A.J., Harvey, A.: JSPAM: A restricted three-body code for simulating interacting galaxies. \emph{Astronomy and Computing}, \textbf{16}, 26--33 (2016).
%
\bibitem{ZS17} Zotos, E.E., Suraj, M.S. Basins of attraction of equilibrium points in the planar circular restricted five-body problem, \emph{Astrophys. Space. Sci.,} \textbf{363}, 20 (2017)
%
\bibitem{Z18} Zotos, E.E.: On the Newton--Raphson basins of convergence of the out-of-plane equilibrium points in the Copenhagen problem with oblate primaries. \emph{Int. J. Non-Linear Mech.}, \textbf{103}, 93--105 (2018)
\bibitem{ZSMA18} Zotos, E.E., Suraj, M.S., Jain, M., Aggarwal, R.: Revealing the Newtonâ€“Raphson basins of convergence in the circular pseudo-Newtonian Sitnikov problem.  \emph{Int. J. Non-Linear Mech.}, \textbf{105}, 43--54 (2018).
%
\end{thebibliography}
\end{document}